\documentclass[twocolumn,aps,prl]{revtex4}
\usepackage{epsfig,amsmath,graphicx}

\newcommand{\PDFfig}[4]{
\begin{figure}
\centering
\includegraphics[width=#4]{#2}
\caption{#3}
\label{#1}
\end{figure}
}

\newcommand{\PDFfigDouble}[4]{
\begin{figure*}
\centering
\includegraphics[width=#4]{#2}
\caption{#3}
\label{#1}
\end{figure*}
}

\begin{document}

\title{Wavelength-Multiplexed Quantum Networks with Ultrafast Frequency Combs}

\author{Jonathan Roslund, Renn\'{e} Medeiros de Ara\'{u}jo, Shifeng Jiang, Claude Fabre and Nicolas Treps}
\affiliation{Laboratoire Kastler Brossel, UPMC Univ. Paris 6, ENS, CNRS; 4 place Jussieu, 75252 Paris, France}

\date{\today}

%
%

\maketitle

Highly entangled quantum networks – cluster states – lie at the heart of recent approaches to quantum computing \cite{Nielsen2006,Lloyd2012}. Yet, the current approach for constructing optical quantum networks does so one node at a time  \cite{Furusawa2008,Furusawa2009,Peng2012}, which lacks scalability. Here we demonstrate the  \emph{single-step} fabrication of a multimode quantum network from the parametric downconversion of femtosecond frequency combs. Ultrafast pulse shaping \cite{weiner2000} is employed to characterize the comb's spectral entanglement  \cite{vanLoock2003}. Each of the 511 possible bipartitions among ten spectral regions is shown to be entangled; furthermore, an eigenmode decomposition reveals that eight independent quantum channels \cite{Braunstein2005} (qumodes) are subsumed within the comb. This multicolor entanglement imports the classical concept of wavelength-division multiplexing (WDM) to the quantum domain by playing upon frequency entanglement as a means to elevate quantum channel capacity. The quantum frequency comb is easily addressable, robust with respect to decoherence, and scalable, which renders it a unique tool for quantum information.

\begin{figure*}
\centering
\includegraphics[width=115mm]{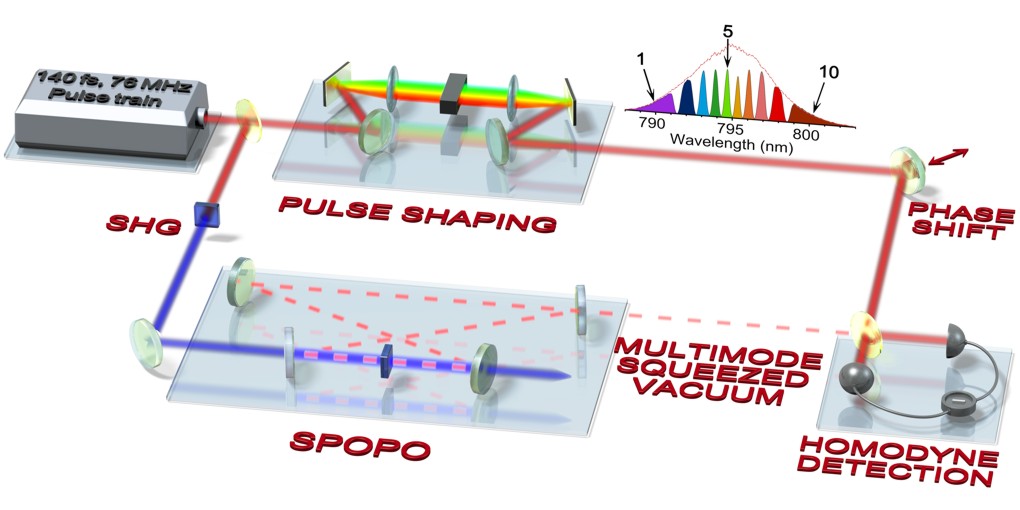}
\caption{Experimental layout for the creation and characterization
of multimode frequency combs. A titanium-sapphire oscillator
produces a $76 \textrm{MHz}$ train of $\sim 140 \textrm{fs}$
pulses centered at 795nm. Its second harmonic synchronously pumps
an OPO, which consists of a 2mm BIBO crystal contained within a
$\sim4 \textrm{m}$ ring cavity. The cavity output is analyzed with
homodyne detection, where the spectral composition of the local
oscillator (LO) is manipulated with a 512-element, programmable 2D
liquid-crystal modulator capable of independent amplitude and
phase modulation \cite{nelson2005}. The spectrum of the LO is
divided into ten discrete bands of equal energy (enumerated on the
figure), and the amplitude and phase of each band may be
individually addressed. By varying the relative phase between the
shaped LO and the SPOPO output, the $\hat{x}$ and $\hat{p}$
quadrature noises of the quantum state projected onto the LO mode
are measured.} \label{fig-experiment}
\end{figure*}


\paragraph*{Theoretical Description}

The use of photonic architectures to realize quantum networks is appealing since photons are immune from environmental disturbances, readily manipulated with classical tools, and subject to high efficiency detection \cite{Pfister2008,Furusawa2011}. We consider here the creation of nonclassical, continuous variable states with an optical parametric oscillator (OPO), in which a pump photon of frequency $2 \omega_{0}$ splits into a pair of lower energy photons subject to energy conservation and the cavity resonance condition.
The generation of a photon pair initiates a nonclassical correlation between the cavity modes $\omega_{-p}$ and $\omega_{p}$, where $\omega_{p} = \omega_{0} + p \cdot \omega_{\textrm{FSR}}$ and $\omega_{\textrm{FSR}}$ is the cavity free spectral range. Given a sufficiently large phase-matching bandwidth, a frequency comb emerges from the cavity with all of the resonant photon pairs independently entangled \cite{pfister2011}.
The inclusion of additional pump photons of frequencies $2 \omega_{0} + p' \cdot \omega_{\textrm{FSR}}$ opens the possibility for richer frequency correlations beyond purely symmetric pair creation.
Femtosecond pulse trains contain upwards of $\sim 10^{5}$
individual frequency modes, and the simultaneous injection of all
these modes into a nonlinear optical element induces an intricate
network of both symmetric and asymmetric frequency correlations
\cite{Pinel2012}. To access such states, a synchronously pumped
optical parametric oscillator (SPOPO), which consists of an OPO
driven by a femtosecond pulse train with a repetition rate
matching the cavity free spectral range, is exploited and creates
correlations governed by the Hamiltonian:

\begin{equation}
\hat{H} = i \hbar  g \sum\nolimits_{m,n}  L_{m,n}\, \hat{a}_{m}^{\dagger} \hat{a}_{n}^{\dagger} + \textrm{h.c.},
\label{hamiltonian}
\end{equation}
where $g$ regulates the overall interaction strength and $\hat{a}_{m}^{\dagger}$ is the photon creation operator associated with a mode of frequency $\omega_{m}$. The coupling strength between modes at frequencies $\omega_m$ and $\omega_n$ is dictated by the matrix $L_{m,n}=f_{m,n} \cdot p_{m+n}$, where $f_{m,n}$ is the phase-matching function \cite{walmsley2001,walmsley2008} and $p_{m,n}$ is the pump spectral amplitude at frequency $\omega_{m}+\omega_{n}$ \cite{patera2010}.

\paragraph*{Frequency Entanglement}

We experimentally demonstrate that the photonic state emerging from a SPOPO pumped below threshold exhibits spectral entanglement across the breadth of the comb.  A femtosecond pulse train is produced with a mode-locked titanium-sapphire oscillator delivering $\sim 140 \textrm{fs}$ pulses, and its second harmonic serves to synchronously pump an OPO as detailed in Fig.~1. Homodyne detection coupled with ultrafast pulse shaping is then employed to fully characterize the quantum properties embedded in the frequency comb.

\begin{figure}
\centering
\includegraphics[width=75mm]{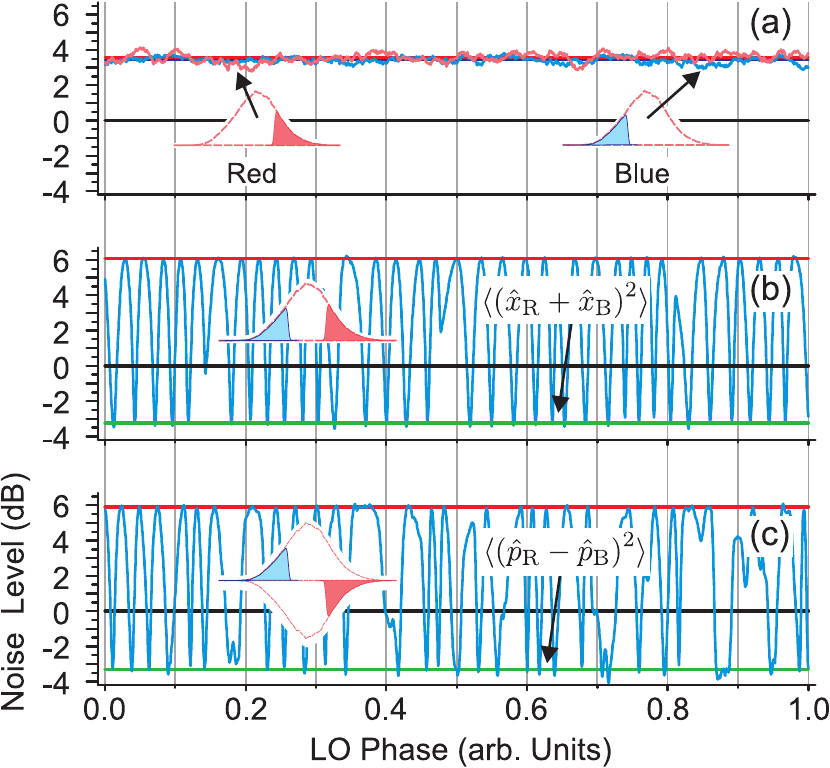}
\caption{(a) Phase-independent, excess noise of $\sim 3.4$dB is
present in both the high (blue) and low (red) frequency bands. (b)
The amplitude of the frequency band sum exhibits $\sim -3.2
\textrm{dB}$ of squeezing. (c) The pulse shaper writes a
$\pi$-phase shift between the spectral wings, and the phase of the
difference also shows a squeezing level of $\sim -3.3
\textrm{dB}$. Hence, the Duan entanglement criterion is readily
verified.} \label{fig-duan}
\end{figure}

The quantum correlations generated in the comb are most easily understood by initially considering only two discrete frequency bands.
For this purpose, amplitude shaping removes all but the low (red) and high (blue) frequencies of the LO spectrum as seen in Fig.~\ref{fig-duan}. The field quadrature components $\hat{x}$ and $\hat{p}$ of these two bands are then measured with homodyne detection while the pulse shaper constructs on-demand entanglement witnesses.
For example, when either frequency band is considered alone, a quadrature-independent excess noise is observed (Fig.~2a) as found with a thermal state. The amplitude of the spectral sum, however, is squeezed (Fig.~2b), which is indicative of a strong intra-comb nonclassical frequency correlation. A $\pi$-phase shift is then applied to the red frequency band with the pulse shaper, which reveals that the phase of the spectral difference is also squeezed (Fig.~2c). With these two measurements, the Duan inseparability criterion \cite{duan2000} $\langle \left( \hat{x}_{R} + \hat{x}_{B} \right)^{2} \rangle + \langle \left( \hat{p}_{R} - \hat{p}_{B} \right)^{2} \rangle = 0.94 \pm 0.03 < 2$ is fulfilled and indicates that the two frequency bands are indeed entangled. In diagnosing these correlations, the pulse shaper has assumed the role of a traditional beam splitter but in a frequency-dependent fashion. We may also go a step further and infer the $\hat{x}$ and $\hat{p}$ conditional variances with the frequency correlation coefficients from Figs.~2b,c. The product of these conditional variances is $\Delta^{2} x_{R|B} \cdot \Delta^{2} p_{R|B} \simeq 0.58 \pm 0.21 < 1$, which also satisfies the more stringent condition for EPR entanglement between the spectral wings of the SPOPO output \cite{bowen2003}.

\begin{figure}
\centering
\includegraphics[width=80mm]{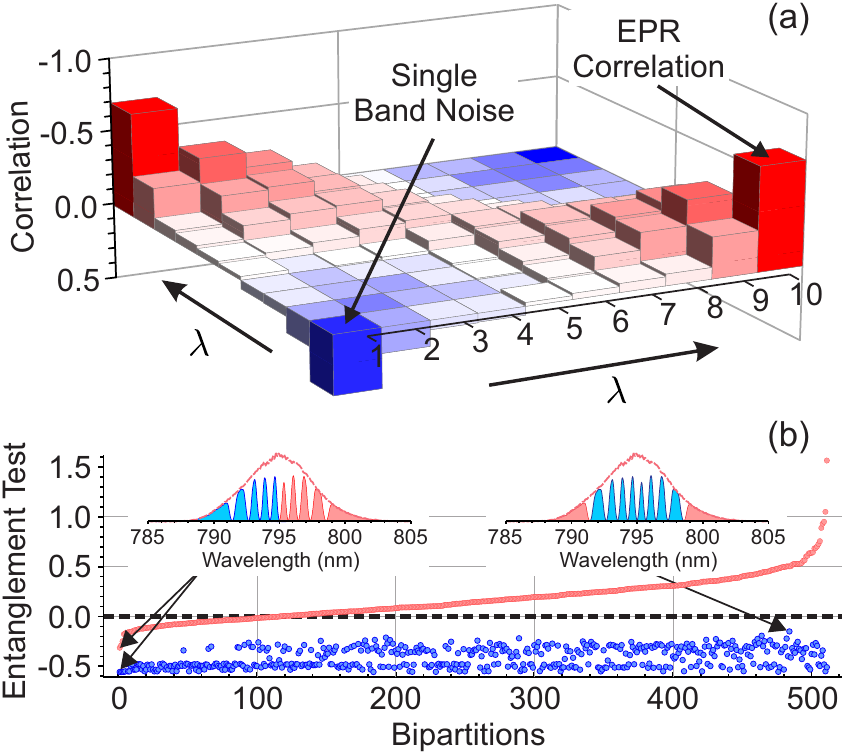}
\caption{(a) Noise correlation matrix, defined as $C^{x}_{i,j} =
\langle x_{i} x_{j} \rangle / \sqrt{\langle x_{i}^2 \rangle
\langle x_{j}^2 \rangle} - \delta_{i,j} \langle x_{vacuum}^2
\rangle / \langle x_{i}^2 \rangle$ for the $x$-quadrature. It is
important to note that the vertical axis is inverted. (b) EPR
(red) and PPT (blue) inseparability criteria for all 511 bipartite
combinations of the 10 spectral bands, ordered according to
increasing EPR values. The black dotted line is taken to be the
entanglement boundary for both tests. All 511 bipartitions possess
a PPT value below the boundary, which indicates complete
non-separability for the state. Additionally, 115 biparitions
satisfy the more stringent criterion for EPR entanglement.}
\label{fig-ppt}
\end{figure}


\paragraph*{Covariance Matrix Measurement}

In order to more aptly characterize the intra-comb entanglement across the entire spectrum, the LO is divided into ten frequency bands of equal energy as seen in Fig.~1, and the $\hat{x}$- and $\hat{p}$-quadrature noises for each spectral region and all possible pairs of regions are determined. The 55 requisite homodyne measurements are acquired in a period of approximately thirty minutes, and allow assembly of the state's full covariance matrix. We have observed that cross-correlations of the form $\langle \hat{x} \, \hat{p} \rangle$ are absent, which permits the covariance matrix to be cast in a block diagonal form. Fluctuations and correlations departing from the vacuum level are shown in Fig.~3a for the $\hat{x}$-quadrature.
A spectrally-dependent distribution of excess noise is evident (diagonal blue peaks) with the preponderance of its occurrence in the spectral wings. Concomitantly, the bulk of the frequency correlation occurs between the two opposing spectral wings (off-diagonal red peaks).

The inseparability of individual frequency bands from the conglomerate structure is probed with the positive partial transpose (PPT) criterion \cite{simon2000}, which is applied to all 511 possible frequency band bipartitions. Every bipartition is entangled, and the degree of its inseparability is assessed by the magnitude of the corresponding Heisenberg inequality violation. As seen in Fig.~3b (blue points), the absence of any partially separable form implies that the SPOPO output constitutes a completely non-separable, genuine 10-partite state \cite{Braunstein2005}.
The bipartitions that induce the largest and smallest physicality violations are of particular interest. Frequency bands diametrically opposed to the central wavelength are the most strongly entangled (Fig.~3b), whereas the partition that disconnects the two spectral wings from the remaining structure is the most weakly entangled. It is important to realize that in the case of a solitary pump frequency, all of the partitions possessing a reflection symmetry about the central wavelength would not be entangled (e.g., the largest PPT value in Fig.~3b). The entanglement of such structures demands asymmetric frequency correlations afforded upon the simultaneous downconversion of multiple pump frequencies.

As before, it is likewise possible to consider EPR entanglement for each of these bipartitions. Upon doing so, 115 frequency bipartitions turn out to additionally satisfy the more stringent condition for EPR entanglement as presented in Fig.~3b (red points).
Analogous to the PPT criterion, the bipartition that displays the strongest EPR entanglement corresponds to bisecting the spectrum at the central wavelength.

\begin{figure}
\centering
\includegraphics[width=85mm]{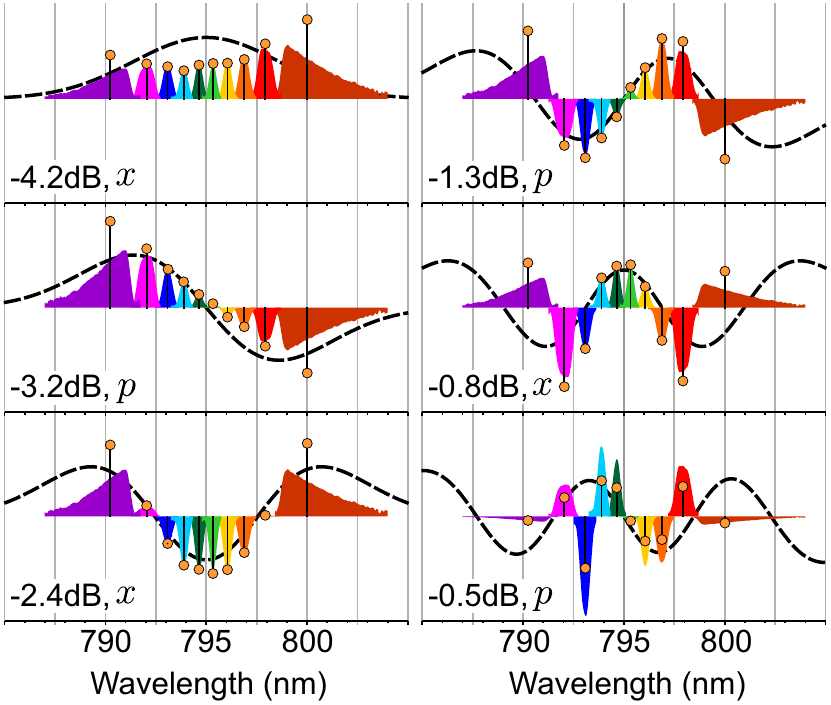}
\caption{(Amplitude spectra and corresponding squeezing values and
quadratures for the leading six experimental supermodes retrieved
from the covariance matrix. The stick spectra represent the
experimentally measured orthonormal vectors from which the
spectral form of the supermodes is constructed. The black trace is
the approximate theoretical Hermite-Gauss form for the supermodes,
where the appropriate spectral width is determined from the SPOPO
above-threshold spectrum.} \label{fig-modes}
\end{figure}


\paragraph*{Modal Decomposition}

While germane for quantum information processing, the multipartite entangled nature of the comb is an extrinsic characteristic as it depends upon a user-specified allocation of individual frequency bands.
For example, although a single mode squeezed beam theoretically acquires multipartite character by simply dividing it with a beamsplitter, the state remains intrinsically monomode \cite{treps2005}. An alternative theoretical description of the state is gleaned upon diagonalizing the interaction matrix $L_{m,n}$ to reveal a frequency-decorrelated modal representation \cite{Leuchs2002,patera2010}. The ensuing basis of ``supermodes'' $\hat{S}_{k}$, which are linear combinations of the original, single frequency modes, permits rewriting the total Hamiltonian as a sum of single-mode squeezing Hamiltonians independently acting on each supermode:

\begin{equation}
\hat{H} = i \hbar  g \sum\nolimits_{k} \Lambda_{k} \, \hat{S}_{k}^{\dagger \, 2}+  \textrm{h.c.},
\label{sqz-hamiltonien}
\end{equation}
where $\Lambda_k$ is the eigenvalue associated with the supermode $\hat{S}_k$.
The eigenspectrum specifies the number of non-vacuum qumodes contained in the SPOPO output and their associated degree of nonclassicality. Thus, the quantum comb is equivalently analyzed either as an entangled state in the multipartite basis exemplified by Fig.~3a, or as a set of uncorrelated squeezed states in a multimode basis \cite{Braunstein2005-irreducible}.

A set of orthonormal experimental supermodes is extracted from the measured covariance matrix
and reveals that 8 squeezed modes are contained within the conglomerate comb structure (see annexe). The spectral composition and squeezing level for the leading six modes are shown in Fig.~4. The squeezing quadrature ($\hat{x}$ or $\hat{p}$) is observed to alternate between successive modes, which is in agreement with theory \cite{patera2010}. Consequently, the SPOPO functions as an \emph{in situ} optical network consisting of an assembly of independent OPOs and phase shifters.

The spectral composition of each experimental mode also displays generally good agreement with Hermite-Gauss polynomial forms, which approximate the predicted supermode structure \cite{patera2010}. The spectral width $\Delta \lambda_{k}$ of these supermodes $\hat{S}_{k}$, however, increases with mode index $k$ as in $\Delta \lambda_{k} =\sqrt{2 k+1} \cdot \Delta \lambda_{0}$, and the diminished overlap with the LO in the spectral wings becomes apparent with each progressive mode. The inability to resolve high-order modes with the fixed bandwidth of the LO accounts for the decrease in observed squeezing levels. The present observation of 8 squeezed modes does not represent an inherent upper limit to the dimensionality of comb states, and with the adoption of broader bandwidth LO pulses, comb states possessing as many as $\sim 100$ significantly squeezed modes are expected \cite{patera2010}.

In order to corroborate these modes, the retrieved structures of the leading four modes in Fig.~4 are written directly onto the pulse shaper. Squeezing is observed for each of these orthogonal modes, albeit in alternating quadratures.
Finally, it is worth noting that while the experimental supermodes of Fig.~4 are represented in the frequency domain, they are equivalently described in the temporal domain. Hence, the entanglement across the entire frequency breadth of the comb visible in Fig.~3a implies a concomitant creation of temporally entangled structures within the ultrafast pulse \cite{Averchenko2011}.

\paragraph*{Discussion}



The ability to independently control the amplitude and phase of each LO spectral element opens the possibility of constructing a basis change that emulates an arbitrary linear optical network \cite{vanLoock2007}. As such, the quantum noise can be measured in any basis, and we can imagine discovering a basis of pulse shapes that infers the presence of cluster states \cite{vanLoock2008}. Additional calculations indicate that cluster states are indeed subsumed within the ultrafast quantum comb, and spectrally-resolved homodyne detection will enable their experimental manipulation.

In conclusion, we have demonstrated ultrafast frequency combs to be a practical, compact source of massively entangled quantum states. The ability to create top-down multipartite entanglement amongst thousands of frequencies with a single nonlinear interaction provides an unprecedented capability. Additional nodes may then be incorporated into the quantum network by simply increasing the number of frequencies participating in the nonlinear interaction, which does not require an expansion of the optical setup.
High dimensional quantum objects provide a means to elevate photonic channel capacities and thereby multiplex the transmission and processing of information. The network of quantum channels present in the ultrafast comb should find numerous applications in quantum metrology and measurement-based quantum computing. Immediate applications involve spectrally-resolved homodyne detection of the quantum comb \cite{Armstrong2012} as a means for implementing quantum measurement protocols and ultraprecise temporal metrology beyond the Heisenberg limit \cite{Treps2008}.
Consequently, highly multimode photonic sources are expected to become a valuable source for realizing cluster states and facilitating fundamental studies of quantum information processing.

\section{Annexe}

The laser source is a titanium-sapphire mode-locked oscillator delivering $\sim 140 \textrm{fs}$ pulses ($\Delta \lambda \simeq 6 \textrm{nm}$ FWHM) centered at $795 \textrm{nm}$ with a repetition rate of $76 \textrm{MHz}$. Its second harmonic at $397 \textrm{nm}$ serves to synchronously pump an OPO, which consists of a 2mm BIBO crystal contained within a 4m long ring cavity exhibiting a finesse of 27 and escape efficiency of 95\%. In this configuration, the OPO reaches its threshold at a mean pump power of $\sim 75 \textrm{mW}$. Below threshold and in the absence of a seed, the device generates squeezed vacuum with a squeezing level of $\sim -6 \textrm{dB}$ (corrected) when projected onto the unshaped LO as seen in Fig.~\ref{fig-fullsqz}. 

\PDFfig{fig-fullsqz}{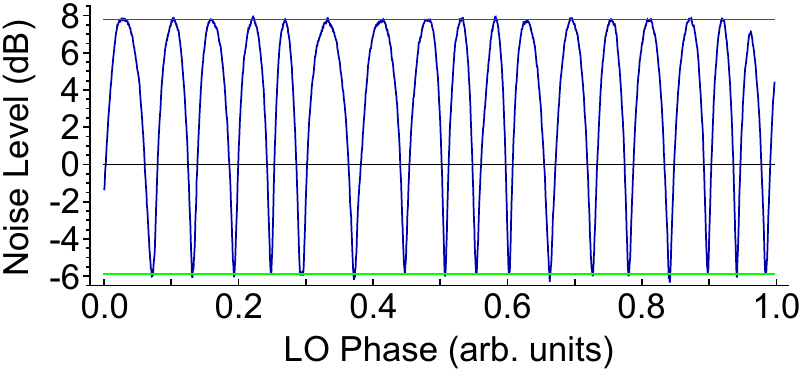}{Noise trace observed with an unshaped LO pulse shape. The state exhibits squeezing and anti-squeezing values of $-5.9 \textrm{dB}$ and $7.8 \textrm{dB}$, respectively.}{80mm}


In order to lock the cavity length to the inter-pulse spacing, a near-infrared beam is phase-modulated at $1.7 \textrm{MHz}$ with an electro-optic modulator (EOM) and injected into the cavity in a direction counter-propagating to the pump and seed. Locking of the cavity is then accomplished with a Pound-Drever-Hall strategy. 

Frequency correlations of the state are investigated with homodyne detection where the local oscillator pulse form is manipulated with a pulse shaper. The $4 f$-configuration shaper is constructed in a reflective geometry with a programmable 512\,x\,512-element liquid-crystal modulator in the Fourier plane.  Application of a periodic spatial grating to the spatial light modulator induces diffraction of the spectrally-dispersed light. The amplitude and phase of the diffracted spectrum are independently controlled by the groove depth and position of the spatial grating, respectively \cite{nelson2005}.

The OPO output is projected by the homodyne detection technique onto the spectrally filtered LO pulse shape to assess the quadrature noise content of the corresponding spectral region. Light detection is performed with silicon photodiodes ($\sim 90\%$ detection efficiency, 100MHz detection bandwidth), and the noise level of the squeezed vacuum is examined at 1MHz. The homodyne visibility is $92 \%$. The cumulative loss of the system is taken to be $\sim 25 \%$ and the measured signals are corrected accordingly. 

\PDFfig{fig-pquad}{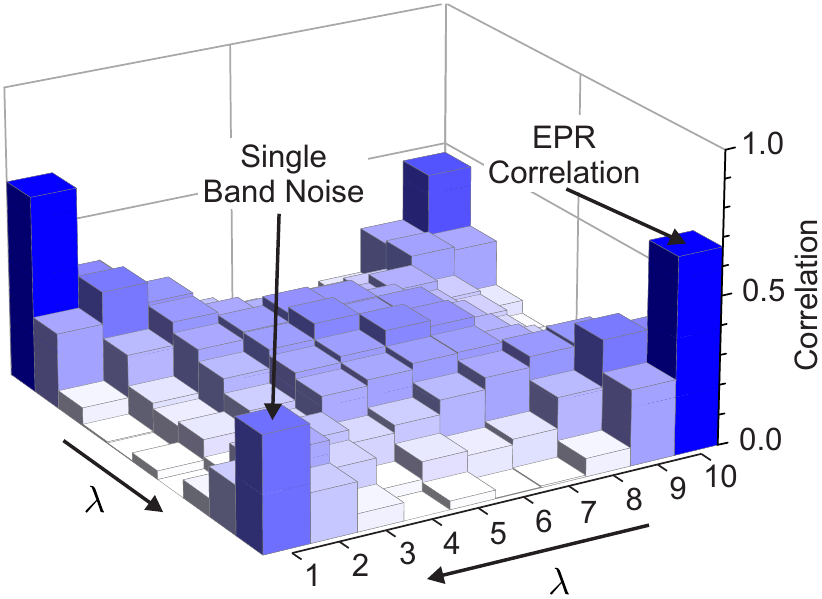}{Noise correlation matrix defined as $C^{p}_{i,j} = \langle p_{i}p_{j} \rangle / \sqrt{\langle p_{i}^2 \rangle \langle p_{j}^2 \rangle} -  \delta_{i,j} \langle p_{i}^2 \rangle_{\textrm{vac}} / \langle p_{i}^2 \rangle$ for the $\hat{p}$ quadrature. The diagonal displays excess noise while the off-diagonal elements demonstrate quantum correlations among disparate spectral regions.}{75mm}

The optical spectrum of the local oscillator is divided into ten frequency bands of equal energy as seen in Fig.~1 of the main text, and spectral holes between the individual regions are imposed to ensure the absence of any spectral overlap. Importantly, the supplemental loss incurred from including these gaps is not accounted for when correcting the noise levels. 
The $\hat{x}$ quadrature is defined as the field quadrature of lowest noise for the unshaped LO pulse. 
The noise dependence on LO phase is then examined for all 55 possible pulse shapes and compared to that of the unshaped LO reference. Each individual trace is observed to follow the phase dependence of the reference, which confirms that the lowest noise level for every spectral combination is present in the $\hat{x}$ quadrature, i.e., there is no rotation of the squeezing ellipse between successive measurements. 


The noise content of both the $\hat{x}$- and $\hat{p}$-quadratures for each spectral region and all possible pairs of regions are measured as shown in Fig.~\ref{fig-sqzTraces}. The individual covariance elements are constructed according to the following equation:

\begin{eqnarray}
\langle x_{i} x_{j} \rangle &=& 
\left[ \langle (x_{i} + x_{j})^2 \rangle - \frac{P_{i}}{P_{i}+P_{j}} \langle x_{i}^2 \rangle - \frac{P_{j}}{P_{i}+P_{j}} \langle x_{j}^2 \rangle \right] \nonumber \\ && \times \, \frac{P_{i}+P_{j}}{2 \sqrt{P_{i} P_{j}}},
\end{eqnarray}
where $P_{i}$ and $P_{j}$ are the optical powers of frequency bands $i$ and $j$, respectively, which are measured with the homodyne photodiodes. 

\textbf{Data Analysis}. Noise fluctuations and correlations departing from the vacuum level for the $\hat{p}$-quadrature are represented in Fig.~\ref{fig-pquad}. Similar to the $\hat{x}$ quadrature, the vast majority of both the excess noise (diagonal peaks) and correlation (off-diagonal peaks) are localized in the spectral wings. As opposed to the $\hat{x}$ quadrature, however, the correlations within the $\hat{p}$ quadrature are positive. 

The purity $\mathcal{P}$, which is an intrinsic property of the state, is accessible from the covariance matrix with the relation $\mathcal{P} = 1 / \sqrt{\textrm{det}(\mathcal{C})}$, where $\mathcal{C}$ is the full covariance matrix of the state \cite{Adesso2007}. This relationship reveals that $\mathcal{P} \simeq 0.45 \pm 0.09$. 
The full covariance matrix is then eigendecomposed, where it is observed that although the individual $\hat{x}$ and $\hat{p}$ block eigenvectors are very similar, they are not identical. 
In general, two matrices are simultaneously diagonalizeable only if they commute. For instance, the commutator of the $\hat{x}$ and $\hat{p}$ covariance matrices for two-mode squeezed vacuum is zero even in the presence of spectrally-uniform loss, and a single eigenvector set perfectly decorrelates both quadratures. The introduction of a spectrally-dependent loss, however, spoils the commutation of the $\hat{x}$ and $\hat{p}$ blocks, and the magnitude of a commutator element is directly proportional to the spectral asymmetry of the loss. The output coupler of the SPOPO does not have a perfectly flat transmission window over the bandwidth of the pulse, and the magnitude of the commutator elements is largest in the spectral wings. Consequently, this asymmetry explains the fact that it is not possible to perfectly remove both the $\hat{x}$ and $\hat{p}$ correlation with a single basis change. 

\begin{table}
\begin{center}
\caption{Squeezing values and uncertainties (linear scale) for the modes retrieved from the twenty dimensional covariance matrix. Eight modes are clearly non-classical while a ninth mode is not included in a conservative estimate. \label{tab-sqz}}
\medskip
{
\begin{tabular}{c|c|c}
\hline
{Mode} & Squeezing Level & Anti-Squeezing Level \\
\hline
1 & $0.38 \pm 0.07$ & $3.86 \pm 0.12$ \\
2 & $0.48 \pm 0.06$ & $3.62 \pm 0.07$ \\
3 & $0.58 \pm 0.07$ & $2.74 \pm 0.10$ \\
4 & $0.74 \pm 0.06$ & $1.96 \pm 0.06$ \\
5 & $0.83 \pm 0.05$ & $1.41 \pm 0.06$ \\
6 & $0.90 \pm 0.03$ & $1.17 \pm 0.04$ \\
7 & $0.93 \pm 0.03$ & $1.11 \pm 0.03$ \\
8 & $0.95 \pm 0.02$ & $1.06 \pm 0.03$ \\
9 & $0.97 \pm 0.02$ & $1.03 \pm 0.03$ \\
10 & $0.98 \pm 0.02$ & $1.00 \pm 0.02$ \\
\end{tabular}
}
\end{center}
\end{table}

Yet, the commutator elements have a small magnitude.
Therefore, as a means of generating a mode set that efficiently removes the correlation of the original frequency bands, the ten most noise-robust eigenmodes (see statistical sampling procedure detailed below) are selected from the total eigenvector set. These vectors, which are drawn from both the $\hat{x}$ and $\hat{p}$ blocks, are orthogonalized with a Gram-Schmidt procedure, and the covariance matrix is subsequently re-expressed in this basis. The resulting matrix is nearly diagonal and contains the squeezing values for each orthogonalized mode on its diagonal. 


The error level associated with each of these squeezing values is subsequently estimated with a stochastic method. 
The data underlying every trace of Fig.~\ref{fig-sqzTraces} is collected for twenty seconds in order to accrue statistics about the individual squeezing and anti-squeezing values. The noise value for a given trace is then drawn from a normal distribution with a mean provided by the red (green) line, which represents the average of all identified peaks (valleys), and a variance specified by the variance of the peak (valley) amplitudes. After randomly sampling from these 110 normal distributions, the corresponding covariance matrix is constructed in the appropriate manner. A collection of $10^{4}$ individual matrices is assembled, and the Gram-Schmidt orthogonalization procedure is repeated for every matrix, which provides both a squeezing spectrum and orthonormal mode set. The mean squeezing spectrum from this procedure is shown in Fig.~\ref{fig-eigentrace} and the individual noise levels are detailed in Table~\ref{tab-sqz}. Likewise, the average mode structures corresponding to the six highest squeezing values are shown in Fig.~4 of the main text.  

\PDFfig{fig-eigentrace}{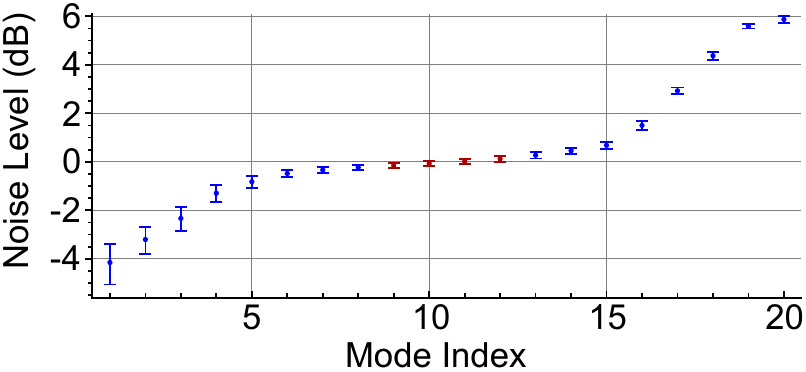}{Mean noise levels and uncertainties (dB) for each of the orthogonalized Gramm-Schmidt modes. The eight modes that are clearly distinguished from the vacuum level are indicated in blue, and modes ascribed to vacuum fluctuations are indicated in red.}{85mm}


\PDFfigDouble{fig-sqzTraces}{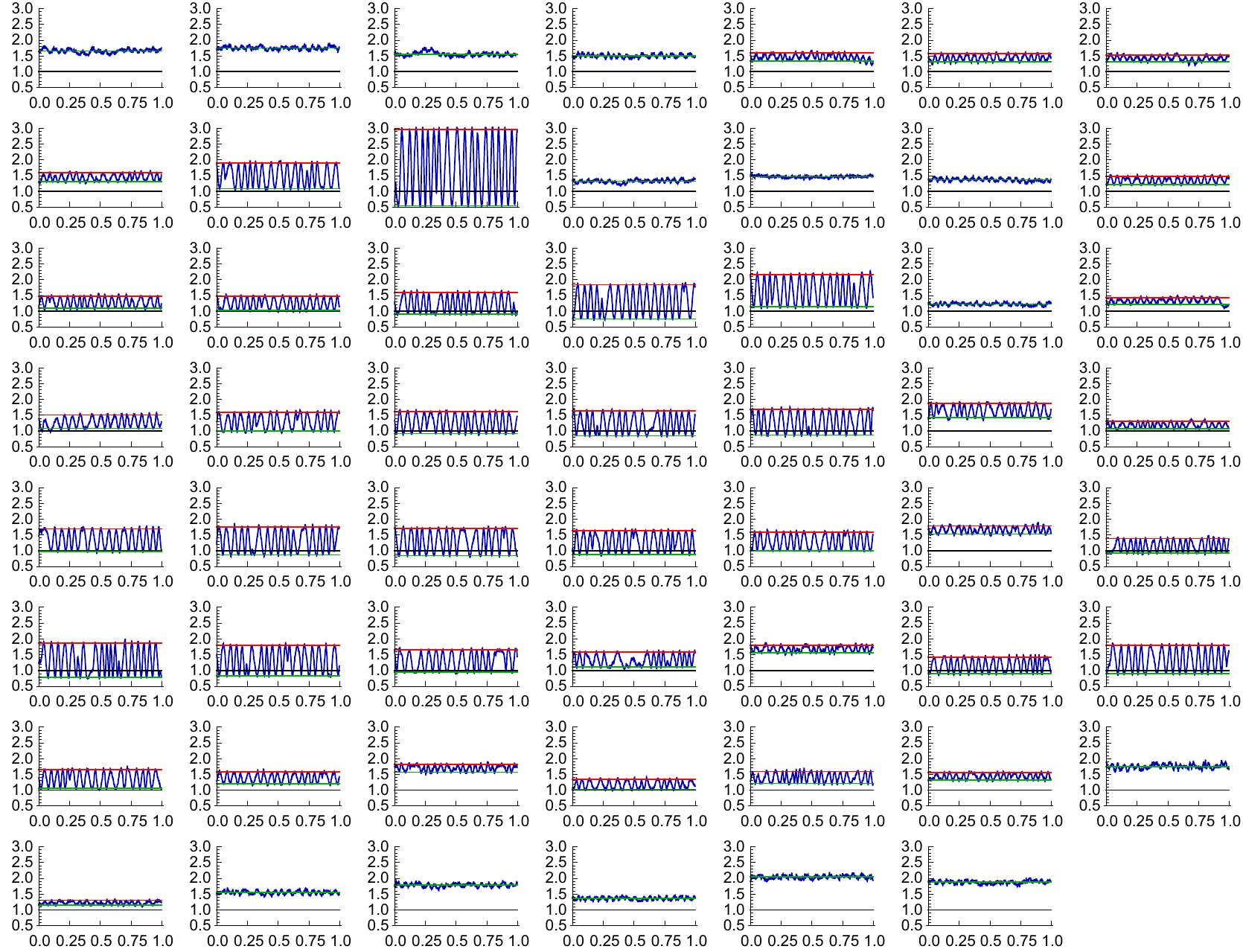}{Squeezing traces for determination of the ten frequency band covariance matrix. The traces are sequentially arranged according to the index scheme $\{i,j\}$, which indicates that the pulse shaper selects frequency bands $i$ and $j$ (e.g., \{1,1\}, \{1,2\},...,\{1,10\},\{2,2\},\{2,3\},...). In each trace, the black line represents the vacuum level while the green and red lines indicate the average squeezing and anti-squeezing levels, respectively. The LO phase is depicted with arbitrary units.}{180mm}

The mean spectrum is generally noise-robust and reveals that eight modes are statistically distinct from the vacuum level. Although a ninth mode may also be considered to exhibit non-classicality, its uncertainty is at the vacuum level, and is therefore not included in a conservative estimate. 


\

The research is supported by the ANR project \emph{Qualitime} and
the ERC starting grant \emph{Frecquam}. C. Fabre is a member of
the Institut Universitaire de France.

\bibliographystyle{apsrev}

\end{document}